\documentclass{raa}            

\usepackage{graphicx,times}             
\usepackage{natbib}
\bibpunct{(}{)}{;}{a}{}{,}

\begin{document}

    \title{UV and NIR size of the HI selected low surface brightness galaxies}

   \volnopage{Vol.0 (20xx) No.0, 000--000}      
   \setcounter{page}{1}          

   \author{Cheng Cheng\inst{1},
        Wei Du\inst{2,3},
        Cong Kevin Xu\inst{1},
        Tianwen Cao\inst{1},
        Hong-Xin Zhang\inst{4,5},
        Jia-Sheng Huang\inst{1},
        Chuan He\inst{1},
        Zijian Li\inst{1},
        Shumei Wu\inst{1},
        Hai Xu\inst{1},
        Y. Sophia Dai\inst{1},
        Xu Shao\inst{1},
        Marat Musin\inst{1}
        }
   \institute{
        Chinese Academy of Sciences South America Center for Astronomy, National Astronomical Observatories, CAS, Beijing 100101, China\\
        \email{chengcheng@nao.cas.cn}
        \and
        National Astronomical Observatories, Chinese Academy of Sciences (NAOC), 20A Datun Road, Chaoyang District, Beijing 100101, China 
        \and
        Key Laboratory of Optical Astronomy, NAOC, 20A Datun Road, Chaoyang District, Beijing 100101, China \\
        \and
        CAS Key Laboratory for Research in Galaxies and Cosmology, Department of Astronomy, University of Science and Technology of China, Hefei, Anhui 230026, People's Republic of China
        \and
        School of Astronomy and Space Science, University of Science and Technology of China, Hefei 230026, People's Republic of China
        }

\vs\no
   {\small Received~~20xx month day; accepted~~20xx~~month day}

\abstract{ 
How does the low surface brightness galaxies (LSBGs) form stars and assemble the stellar mass is one of the most important questions to understand the LSBG population. We select a sample of 381 HI bright LSBGs with both Far Ultraviolet (FUV) and Near Infrared (NIR) observation to investigate the star formation rate (SFR) and stellar mass scales, and the growth mode. We measure the UV and NIR radius of our sample, which represent the star-forming and stellar mass distribution scales. We also compare the UV and H band radius-stellar mass relation with the archive data, to identify the SFR and stellar mass structure difference between the LSBG population and other galaxies. Since galaxy HI mass has a tight correlation with the HI radius, we can also compare the HI and UV radii to understand the distribution of the HI gas and star formation activities. Our results show that most of the HI selected LSBGs have extended star formation structure. The stellar mass distribution of LSBGs may have a similar structure as the disk galaxies at the same stellar mass bins, while the star-forming activity of LSBGs happens at a larger radius than the high surface density galaxies, which may help to select the LSBG sample from the wide-field deep u band image survey. The HI also distributed at a larger radius, implying a steeper (or no) Kennicutt-Schmidt relation for LSBGs.
\keywords{Galaxies: evolution -- Galaxies: dwarf -- Ultraviolet: galaxies -- Galaxies: star formation}
}

\titlerunning{UV and NIR size of LSBGs}
\authorrunning{Cheng et al.}
   \maketitle

%
%
\section{Introduction}
Low surface brightness galaxies (LSBGs) draw more and more attention in studies of the galaxy formation and evolution. It is not only because they are mainly young galaxies with faint and diffuse nature, which might provide clues to the galaxy formation, but also because of their interesting connection with their dark matter halos \citep{1997ARA&A..35..267I, 1997PASP..109..745B}.

Traditionally, LSBGs are selected by the centre surface brightness of a galaxy in B band \citep{1996MNRAS.280..337M}. Since B band is more sensitive to the star formation, such selection of the LSBG sample may cause a bias for the galaxy population with a diffuse star formation structure. Moreover, an average LSBG has a relatively young stellar population and low metallicity \citep{2017ApJ...837..152D}. The typical stellar mass is much lower than the $M^*$ at the similar redshift \citep{Du2020}, indicating an early forming stage of the galaxies.

However, due to the low surface brightness nature, it is difficult to obtain the optical spectrum with enough signal noise ratio even to identify the redshifts \citep{2015ApJ...798L..45V, 2018ApJ...857..104G, 2018ApJ...866..112G}. Thus a large sample of the LSBGs with reliable redshifts is still quite scant, limiting the statistical study of LSBGs. Luckily, most LSBGs are low mass galaxies, which tend to be rich in HI \citep{Huang2012, Maddox2015}. So the wide field HI survey projects such as ALFALFA \citep{2011AJ....142..170H, 2018ApJ...861...49H}, combined with SDSS optical images, would provide us optimal means to obtain a large sample of LSBGs with reliable HI redshift. Our previous series study based on the LSBGs selected from ALFALFA sample with optical surface brightness measured from SDSS images \citep{Du2015, 2019ApJ...880...30H, Du2019} have shown that the LSBGs have much lower metallicity \citep{2017ApJ...837..152D}, follow the Tully-Fisher relation \citep{Du2019}, and have similar mass-light ratios to the high surface brightness galaxies \citep{Du2020}. The H$\alpha$ images of our HI selected LSBG sample also shows that the star formation surface density has a very weak correlation with the gas surface density, the star formation surface density is much lower than the prediction of Kennicutt-Schmidt law \citep{Kennicutt1998, Kennicutt2012}, leading to a significantly longer gas depletion time scales, and low star formation efficiency \citep{Lei2018, Lei2019}.

Although a typical LSBG is rich in neutral hydrogen gas, the star formation rate (SFR) is not high. For one thing, LSBGs locate in the lower stellar mass end of the star formation galaxy main sequence (MS), where the SFR is lower.
For the other, cold H$_2$ gas, which is the direct fuel of the star formation, is rarely detected in LSBGs. Moreover, the transition from HI to H$_2$ may be also very low efficient in low metallicity environment\citep{2005ApJ...626..627O, 2012MNRAS.426..377G}. H$_2$ gas forms on the surface of dust grains \citep{1979ApJS...41..555H} and requires high gas density \citep{Leroy2008}. But LSBGs have short star formation history, thus may not be able to accumulate enough dust. The extended morphology also implies the HI gas are not concentrated in the galaxy centre \citep{2017AJ....154..116C}. 

The study of the star formation in LSBG is crucially important to the understanding of the formation process in the young galaxy population with extreme properties such as low mass, low dust and low metallicity. One simple and yet fundamental parameter to show the galaxy formation and evolution path is the galaxy size in different bands \citep{Cheng2020}. Previous studies show that massive star-forming galaxies have a larger star formation radius than that of the stellar distribution radius, implying an ``inside-out'' growth mode because most of the stars have concentrated in the galaxy centre region, while the star formation is more extended \citep{2012MNRAS.421.1007K}. Since the NIR bands trace the flux from the old stellar population, which posses the majority of the stellar mass, and can be used to trace the stellar mass distribution, especially for low mass galaxies \citep{2019ApJ...877..103S, 2019ApJ...885L..22S}. On the other hand, SFR radius can be derived from several spatially resolved indicators originated from the young stellar population emission, or re-radiation, such as the UV, FIR broadband images, the H$\alpha$ map from narrowband images or IFU observations. Each SFR indicators has its advantage, but most of them are suffered from the dust extinction issue. Since the LSBGs have low dust abundance \citep[e.g., the g - r colour in ][]{Du2015}, UV images might be the simplest method to reveal the SFR distribution and to estimate the star formation radius.

By comparing the UV and NIR bands HST images in the GOODS-North field, we have found the low mass star-forming galaxies can have an ``outside-in'' growth mode, which means that the star formation is active in the galaxy centre, while the stars are already formed in a larger radius \citep{Cheng2020}. Nevertheless, our previous results based on the HST images only covered the UV bright galaxy with optical spectroscopy redshift, which usually have a high centre surface brightness. LSBGs as an influential member of low mass galaxy population is absent in \citet{Cheng2020}. So in this work, we select a sample of HI bright LSBGs with GALEX and UKIRT detection, aiming to study the growth mode of the LSBGs.

Throughout this paper, we assume a standard $\Lambda$CDM cosmology with $H_0=70\, \rm km/s/Mpc$, $\Omega_{\rm M} = 0.3$, and $\Omega_{\rm \Lambda} = 0.7$. All the magnitude are in the AB magnitude system \citep{1983ApJ...266..713O}.

\section{Sample selection}

Our previous work \citep{Du2015} has built a sample of 1129 LSBG with HI detected by Arecibo telescope, which is the parent sample of this work. The HI selection enables us to have the spec-z of the galaxies fainter than the SDSS spec-z availability, and reach to a low stellar mass galaxy population. We select the LSBGs based on the classical criteria that B band center surface brightness $\mu_{0, \rm B} > 22.5 \rm mag/arcsec^2$. The galaxy centre brightness is derived from the SDSS g, r band images by GALFIT \citep{2002AJ....124..266P, 2010AJ....139.2097P} and calibrated to the B band \citep{2002AJ....123.2121S}. The follow up multi-wavelength study show that 544 LSBGs in our sample have both GALEX and UKIRT observations \citep{Du2020}. We further remove the galaxies that have the bright neighbour within 2$\times R_{\rm optical}$, and remove the galaxies at the edge of the GALEX or UKIRT images. We also remove the galaxy with no GALEX UV detections because this work aims to focus on the properties of the star-forming LSBGs. Our final sample contains 381 galaxies. The stellar mass measured from multi-wavelength catalogue has been present in \citet{Du2020}. We adopt the galaxy distance given by ALFALFA catalogue.

LSBGs are mainly a population of low mass galaxies. To understand the LSBG properties in a wider picture, we compare the LSBG sample with the low mass galaxy sample in \citet{Cheng2020}, which includes the galaxies at $0.05<z_{\rm spec}<0.3, F606W < 24$ AB mag in the Cosmic Assembly Near-infrared Deep Extragalactic Legacy Survey \citep[CANDELS, ][]{2011ApJS..197...35G, 2011ApJS..197...36K} GOODS-North field \citep{2019ApJS..243...22B} with a stellar-mass range $10^7<M_*/M_{\odot}<10^{11.4}$. The CANDELS GOOD-North field has been covered by {\it Hubble} Deep UV Legacy Survey \citep[HDUV, ][]{2018ApJS..237...12O} in F275W, F336W band. The high spatial resolution of this low-z low mass sample should include more high surface brightness galaxies, which are easier to measure the spec-z \citep[for more details, see][]{Cheng2020}. We also compare our sample with the low redshift spatially resolved galaxy sample in \citet{Trujillo2020}. This sample includes 1005 galaxies with a stellar-mass range that $10^7<M_*/M_\odot<10^{12}$ at $z<0.09$, including various galaxy morphology. The multi-wavelength spatial resolution and the large stellar mass range enable us to compare the LSBG H band radius and analyze the possible origin of the offsets (if any) between the samples. To understand the stellar mass, radius relation of LSBG and other stellar systems, we also compare our results with \citet{Misgeld2011}, including the stellar systems in 10 orders of magnitude in mass. Recent results of the UDS sample are also included in this work \citep{2015ApJ...798L..45V, 2020arXiv200205171B}.

\section{Data reduction}

Resolution of the GALEX image is about 5 arcsec. The star formation regions in the galaxy disk are clumpy, so the UV images are not as smooth as the optical images with the current resolution. Since we only need a rough UV scale estimation, we convolve the UV image with a Gaussian kernel that declines the UV image spatial resolution into 7 arcsec, 10 arcsec and 15 arcsec FWHM. Then we measure the half-light radius with SExtractor individually 
\footnote{GALEX images of our LSBG sample are very clumpy, so the SExtractor would treat the clumps as several targets. We decline the image resolution so that the SExtractor can treat the galaxy as one target.}. 
We add 1 $\sigma$ noise to the UV image and measure the half-light radius again to estimate the uncertainty. Since the convolution would smooth the noise and give a smaller uncertainty of the radius, we compare the radius measured from different kernel convolved image and conclude the scatter of the UV radius is about 0.05 kpc.

\begin{figure}
    \centering
    \includegraphics[width=0.8\textwidth]{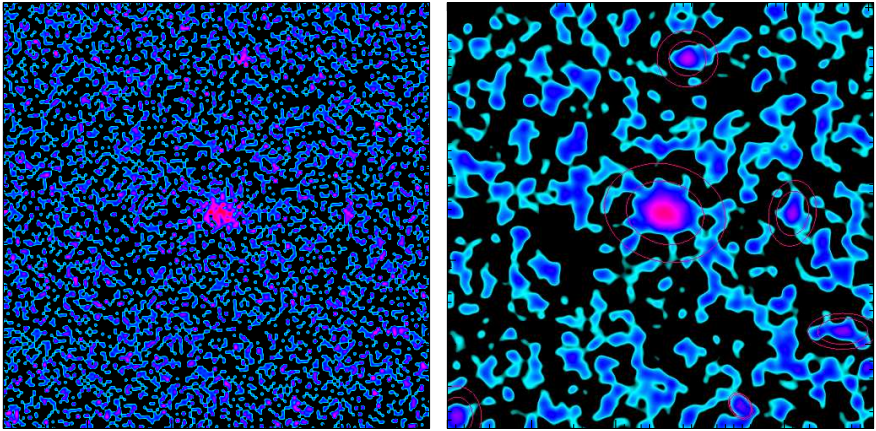}
    \caption{Example of the radius measurements in UV images in the scale of $200'' \times 200''$. The left panel shows one example of LSBG, which shows a clumpy morphology. We convolve the left panel image to a resolution of 7 arcsec in the right panel. Then the morphology of this galaxy is more concentrated to the centre. Then we use SExtractor to measure the half-light radius in the convolved image, and de-convolve the radius by $r_{\rm intrinsic}^2 = r_{\rm measure}^2 - r_{\rm PSF}^2$. Galaxies with a measured radius larger than $r_{\rm PSF}$ are treated as un-resolved targets and will only give upperlimits.}
    \label{Example-UV}
\end{figure}

As a consistency check, we also derive the growth curve of the Gaussian kernel convolved UV image, which is the aperture flux from a series of the radius with the centre given by SExtractor. These two methods give consistent results. But since large aperture photometry (at least 5'' for UV images) would include more noise, and the background for aperture photometry within the large aperture is not uniform, aperture photometry would under the uncertainty caused by the large aperture background estimation. Photometry by SExtractor can estimate the large scale background based on the whole image, and thus would provide more accurate photometry results and better half-light radius results. 

Star formation in galaxy happens discretely The UV morphology would be very clumpy, which would mislead the source extraction and radius measurements. We convolve this kind of image by the kernel larger than the GALEX PSF, which concentrate the UV emission and smooth the UV flux. However, the convolution would involve the noise from the nearby source. Moreover, for the large galaxies with clumpy UV morphology, it is also hard to identify whether the clumps at large radius also part of the central galaxy, or object at the line of sight. In this work, the LSBG UV images are extended and clumpy. To minimize the effect on the radius measurement, we select our sample with no nearby galaxies within 2 times the optical radius. On the other hand, for the LSBG with very clumpy and extended UV morphology should have a larger half-light radius, and thus a large star formation size, which does not change the main conclusion of this work.

\begin{figure}
    \centering
    \includegraphics[width=0.8\textwidth]{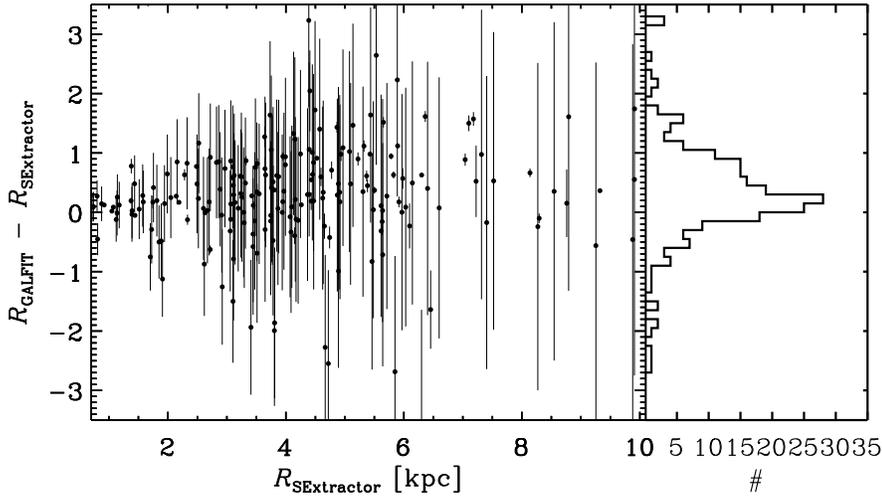}
    \caption{Distribution of the UV half light radius difference between $R_{\rm SExtractor}$ from the non-parameter method (SExtractor) and $R_{\rm GALFIT}$ from the parameter method (GALFIT). The two methods show a consistent results, with a typical scatter about 0.5 kpc.}
    \label{UVcompare}
\end{figure}

Besides the above non-parameter method, we also fit the GALEX UV images with single Sersic function to measure the half-light radius. We adopt the r band galaxy image centre RA, Dec as the centre of the UV image, which will help us to remove the uncertainty of UV image centre that caused by the UV clumps. We fit the UV images by GALFIT \citep{2002AJ....124..266P, 2010AJ....139.2097P}, which is proved to be one of the standard tools to analyse the galaxy morphology. We show the UV half-light radius derived from GALFIT and the direct measurement results in Fig. \ref{UVcompare}. The two radii show good consistency, with a typical scatter about 0.5 kpcs.

\begin{figure}
    \centering
    \includegraphics[width=0.48\textwidth]{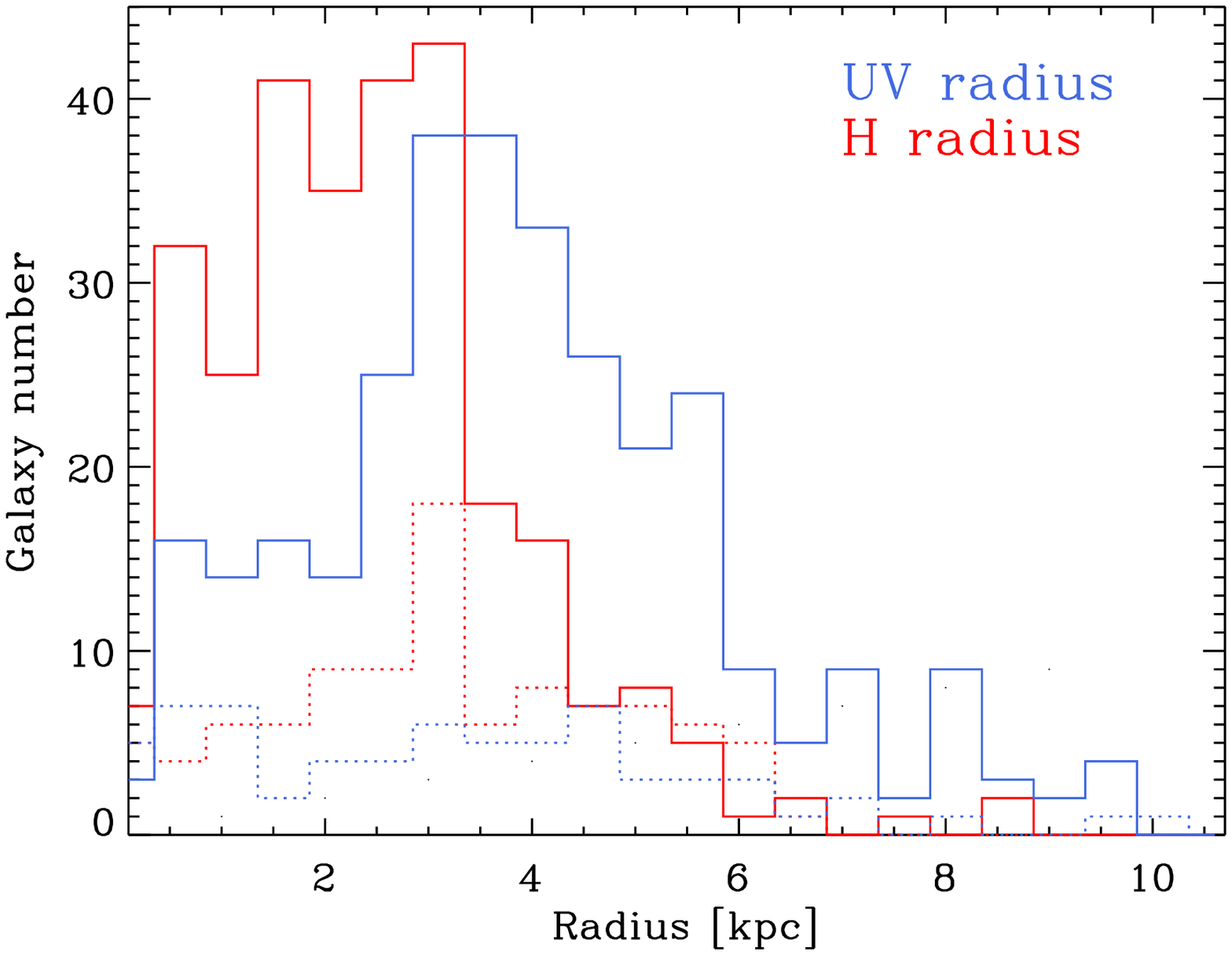}
    \includegraphics[width=0.48\textwidth]{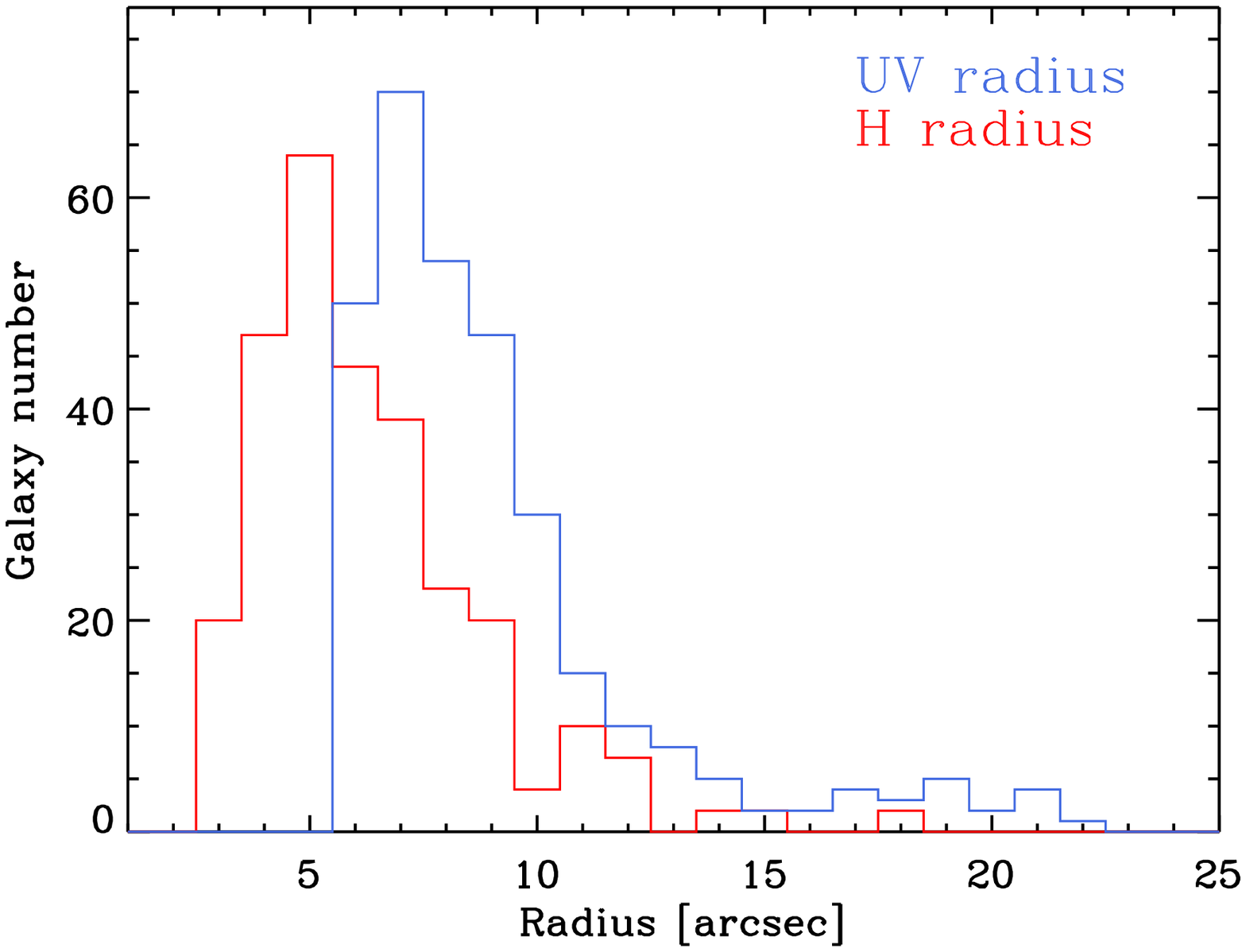}
    \caption{{\bf Left panel: }Histogram of the radius measurement results We show the UV radius in blue colour and the H band radius in red colour. The dot lines show the histogram of the radius upper limits for the targets not resolved in UV or H band images.
    {\bf Right panel: } Histogram of the de-convolved radius in the unit of arcsec. 
    So most of our targets have the size larger than the image resolution.
    }
    \label{hist_radius}
\end{figure}

We use the SExtractor to estimate the half-light radius. We de-convolve the half-light radius by the relation:  $R_{\rm half-light}^2 = R_{\rm measured-size}^2 - R_{\rm PSF-size}^2$. We denote the galaxies with $R_{\rm measured-size} < R_{\rm PSF-size}$ as un-resolved, then we use the $R_{\rm PSF-size}$ as the upper limits of the targets. Fig. \ref{Example-UV} shows one example of the measurement process. The left panel is the UV image while the right panel shows the UV image that convolved into 7'' resolution. The UKIRT H band image has a resolution about 1 arcsec, which is much lower than the UV image. We convolve the H band image into 3 to 7 arcsec FWHM, and measure the H band half-light radius following the same method as the UV radius measurement. We get 312 LSBGs that resolved in UV images; 284 LSBGs that resolved in H band images. There are 253 LSBGs have been resolved in both UV and H bands.

Fig. \ref{hist_radius} shows the histogram for the radius of our targets in UV and H bands. The UV radius of our sample is a bit larger than the H band radius. The radius upper limits of the targets that un-resolved in the images are shown in dot lines. The upper limit size of the LSBGs has a wide range in the unit of kpc.

We fit the SED by fastpp\footnote{https://github.com/cschreib/fastpp} \citet{2009ApJ...700..221K} with the distance given by ALFALFA, and the SED produced in \citet{Du2020}. We employ the \citet{2003MNRAS.344.1000B} stellar population synthesis models (BC03), and choose the Chabrier IMF \citep{2003PASP..115..763C}. We also adopt the \citet{2000ApJ...533..682C} dust extinction law with the attenuation in the range of 0 < Av < 3. The uncertainty of the stellar mass fitting results is about 0.5 dex.

\begin{figure}
    \centering
    \includegraphics[width=0.95\textwidth]{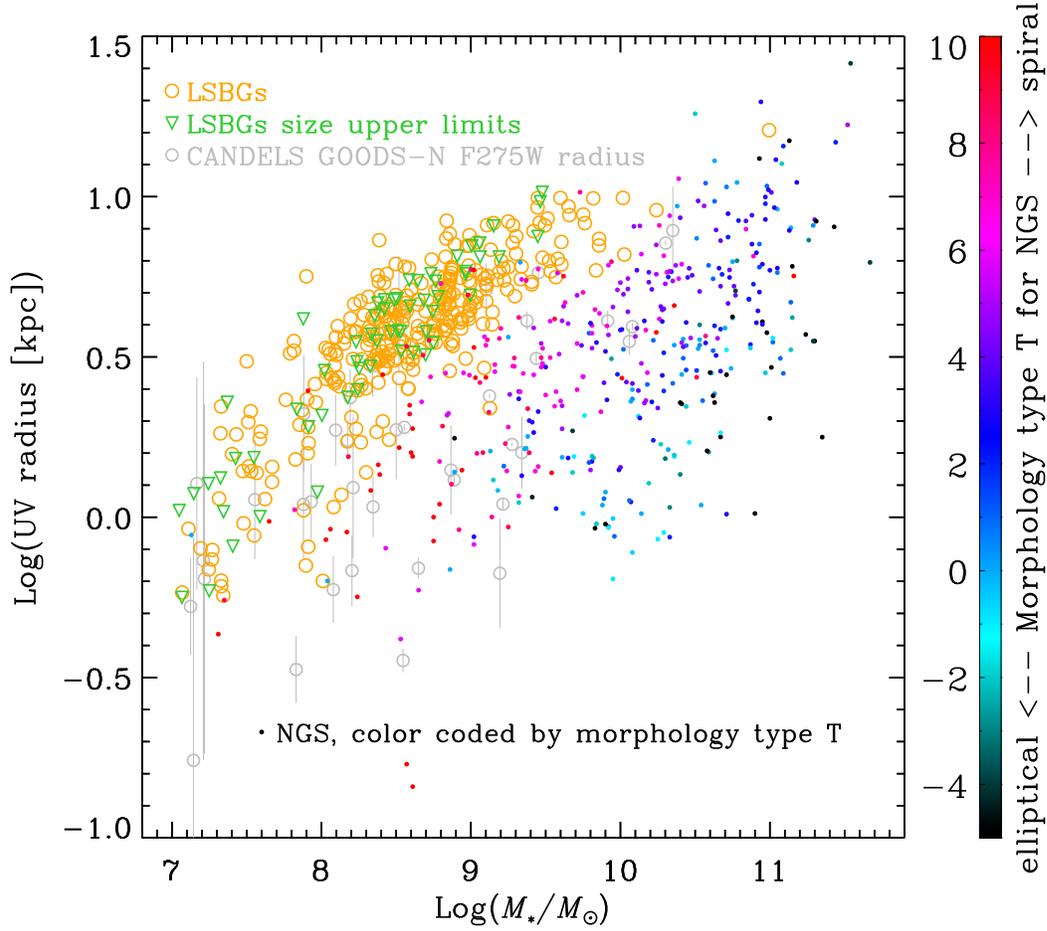}
    \caption{The UV radius v.s. the stellar mass of this LSBG sample (orange open circles for UV resolved sample and green triangle for the size upper limits of the un-resolved UV sample) and the GALEX Nearby Galaxies Survey \citep[NGS, at z $\sim$0 ][]{2007ApJS..173..185G} results (filled circles, colour coded by the morphology type). We also show the HST F275W band half-light radius galaxies at $0.05<Z<0.3$ from CANDELS GOODS-N field \citep{Cheng2020} (grey open circles). Our LSBG sample UV radii are 0.5 dex larger than that of the CANDELS sample, and have a similar UV radius as the local galaxies, but lower stellar mass.}
    \label{UVsize}
\end{figure}
\begin{figure}
    \centering
    \includegraphics[width=0.95\textwidth]{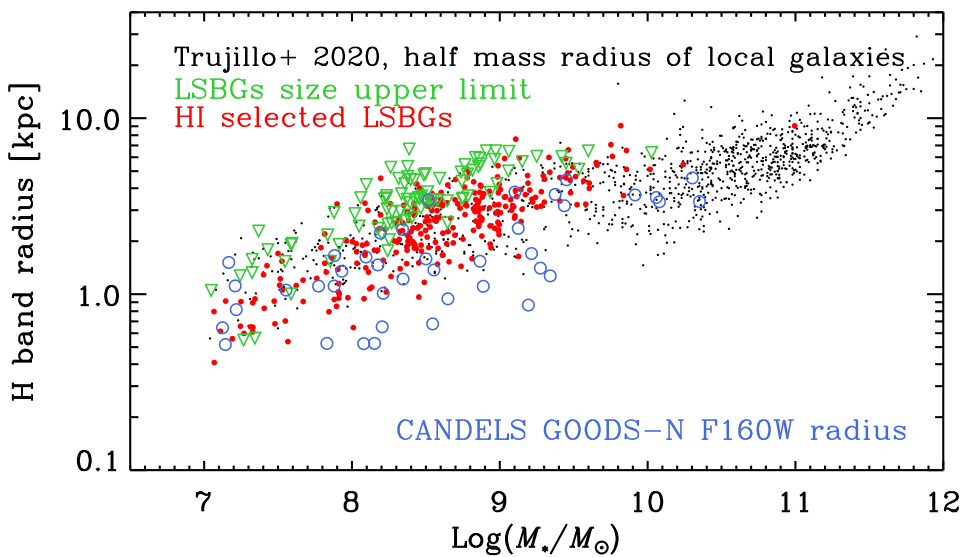}
    \caption{The H band mass size relation of our LSBG sample. We also show the stellar mass v.s. half-mass radius of 1005 low redshift ($z<0.09$) galaxies selected from SDSS as a comparison \citep{Trujillo2020}. We also show the CANDELS results from \citep{Cheng2020} in blue open circles. CANDELS images enable us to measure the radius of the low mass galaxies that not spatially resolved in SDSS images.}
    \label{Hbandsize}
\end{figure}

\section{Results}

We compare the UV band radius of our LSBG sample with the galaxy UV size from GALEX Nearby Galaxies Survey \citep[NGS, at z $\sim$0 ][]{2007ApJS..173..185G} with the latest stellar mass measurements from the z0MGS project \citep{2019ApJS..244...24L} in Fig. \ref{UVsize}. We also show the  CANDELS GOODS-North F275W band half-light radius from \citet{Cheng2020}. The LSBG show a higher UV radius, indicating an extended star formation region than most of the local UV bright galaxies with the same stellar mass range. We also colour the NGS sample with the morphology type T to show the morphology trend and discuss the UV size in Sec. \ref{massUVsize}.

The H band size v.s. stellar-mass relation is shown in Fig. \ref{Hbandsize}. Since the half mass radius and half-light radius ratio for the galaxies about $10^9 M_\odot$ stellar mass is about $0.8-1$ \citep[e.g., the Fig. 1 in ][]{2019ApJ...885L..22S}, we can compare our half-light radius of the LSBGs in H band with the previous half mass radius results in \citet{Trujillo2020}, which includes a sample of 1005 galaxies spans five orders of magnitude in stellar mass ($10^7<M_*/M_\odot<10^{12}$) at a similar redshift range as our LSBG sample ($z<0.09$). Although the mass size relation has different branches for elliptical and disk galaxies \citep{2014ApJ...788...28V}, the LSBG sample mainly has the stellar mass about $10^{7.5}-10^{9.5}M_\odot$, which is the stellar mass range of the local dwarf and disk galaxies in \citet{Trujillo2020}. Our LSBGs H band radii as a low mass disk galaxy sample \citep{Du2019} show a consistent mass size distribution with the results of \citet{Trujillo2020}. 

We also show the CANDELS W160W band radius from \citet{Cheng2020} in Fig. \ref{Hbandsize} by blue open circles. CANDELS results have one order of magnitude higher spatial resolution than the ground-based telescope images and much deeper image limit magnitude. Therefore, the CANDELS survey can enable us to measure the radius of the low mass galaxies that not resolved in the SDSS image. The H band size-mass relation of the CANDELS results indeed show a larger scatter with lower H band radius, implying that the \citep{Trujillo2020} and our LSBGs H band size mass relation may bias to the spatially resolved low mass galaxies.

\begin{figure}
    \centering
    \includegraphics[width=0.95\textwidth]{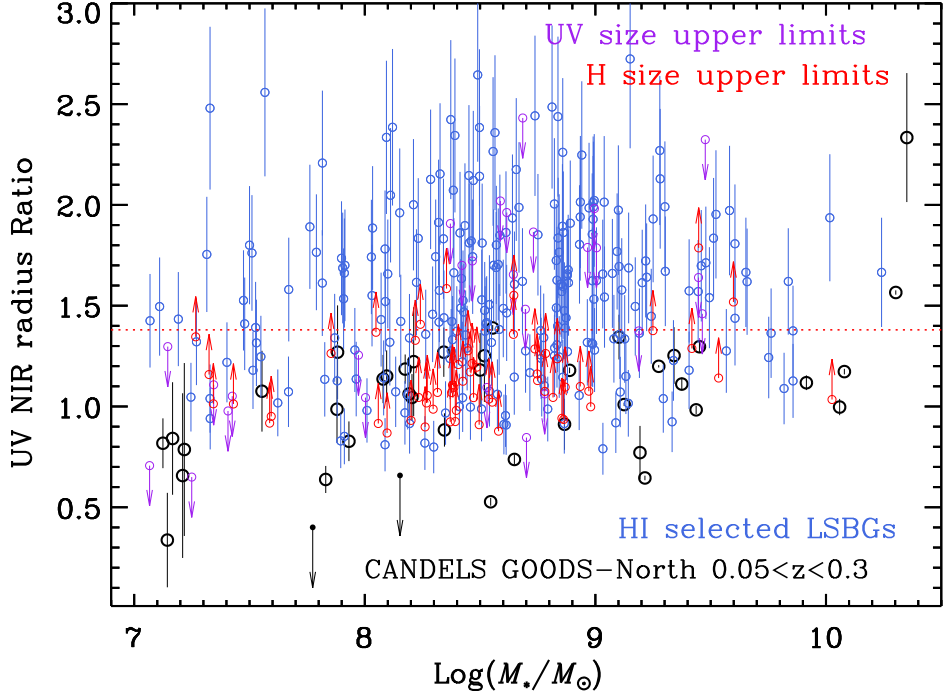}
    \caption{The UV, NIR radius ratio as a function of the stellar mass for the LSBG sample (blue open circles). We show the upper and lower limits of the size ratio for the UV or H band un-resolved LSBGs. As a comparison, we show the UV-NIR radius ratio of the CANDELS sample with open circles, taken from \citet{Cheng2020}. The green dot line shows the results ratio of from GAMA results, where the UV radius is extrapolated from the radius from u, g, r, i, z, J, H, K band average radius. The ratios of LSBGs are consistent with the expectation from the SDSS galaxies results.}
    \label{sizeratio}
\end{figure}

We show the radius of the H and UV band LSBGs in Fig. \ref{sizeratio}. The green dot line is the UV and H band radius ratio based on the average radius in different band \citep{2012MNRAS.421.1007K}. We show our LSBG size ratio as blue circles. For the LSBGs only resolved in UV band image (58 galaxies) or H band images (27 galaxies), we show the size ratio as upper and lower limits in Fig. \ref{sizeratio}. 33 galaxies with neither UV nor H band resolved image are not shown in Fig. \ref{sizeratio}. We can see that majority of the LSBGs have a relatively larger UV size, yielding the on-going star formation occur at larger galaxy radius. We will discuss the affection of the upper limits to our conclusion in Sec. \ref{limits}.

Our LSBGs sample mainly has a larger UV size than the H band with a large scatter at a stellar-mass range about $10^{8.5}M_\odot$. Even in the stellar mass range about $10^7M_\odot$, where our previous work shows that the galaxy may have an ``outside-in'' growth mode, the LSBGs still have a more extended star formation size. The UV and H band size of LSBGs indicate that the LSBGs have a similar stellar mass distribution as the other low mass galaxies, yet the on-going star formation is more extended.

For an exponential disk, the center surface brightness ($\mu_0$) and the efficient brightness ($\mu_{\rm eff} = {\rm mag} + 2.5 \log(\pi r_{\rm eff})$) has a relation that $\mu_0 = \mu_{\rm eff} - 1.83$ \citep{2018ApJ...857..104G}. So the LSBGs selected should have a larger B band radius, yielding a larger radius in the blue band. 

The tight correlation between the HI gas mass and size can help us to estimate the HI radius \citep{2016MNRAS.460.2143W, 2019MNRAS.490...96S}. The HI mass of our LSBG sample is derived from $M_{\rm HI} = 2.35\times 10^5 D^2 F_{\rm HI}$, where the $D$ is the target distance in the unit of Mpc, $F_{\rm HI}$ is the HI flux in the unit of Jy km/s. We estimate the HI size by log(Diameter) = $(\log(M_{\rm HI}) - 6.54)/1.95$ \citep{2013pss6.book...91G}. Fig. \ref{HIsize} show the result of HI and H band size as a function of UV radius. The HI size is typically 3 times larger than the UV radius, while the H band radius is about half the UV radius.

\section{discussion}

We discuss the bias of our sample selection, radius of the LSBG population, and the possible origin of the LSBG size in this section.

\subsection{Bias of our sample selection}

There are in total of 544 targets with both GALEX and NIR observation coverage. To have a better measurement, we remove the 154 galaxies with close counterparts or the galaxies locating in the edge of either GALEX or NIR images, which would not bias our main results. We also remove the 9 galaxies with both GALEX and NIR observations, but have non-detection ($5\sigma$) in either band. The non-detection may be caused by the dust extinction in UV bands, or the shallow survey depth. Since the whole sample we used in this work contains 381 galaxies, much larger than the amount of the non-detection galaxies, we conclude that the removal of the non-detections would not affect our results.

\subsection{Affection of the size limits}\label{limits}

Our results show that most of the HI-selected LSBGs have a more extended UV size than the H band. This conclusion still holds for the H band un-resolved galaxies. For the LSBGs in our sample with UV un-resolved image (in total 60 galaxies) would be the candidate of the LSBGs compact star formation cores \citep{Cheng2020}, and still needs more data to constrain the star formation region (e.g., H$\alpha$ narrow band observation).

\begin{figure}
    \centering
    \includegraphics[width=0.99\textwidth]{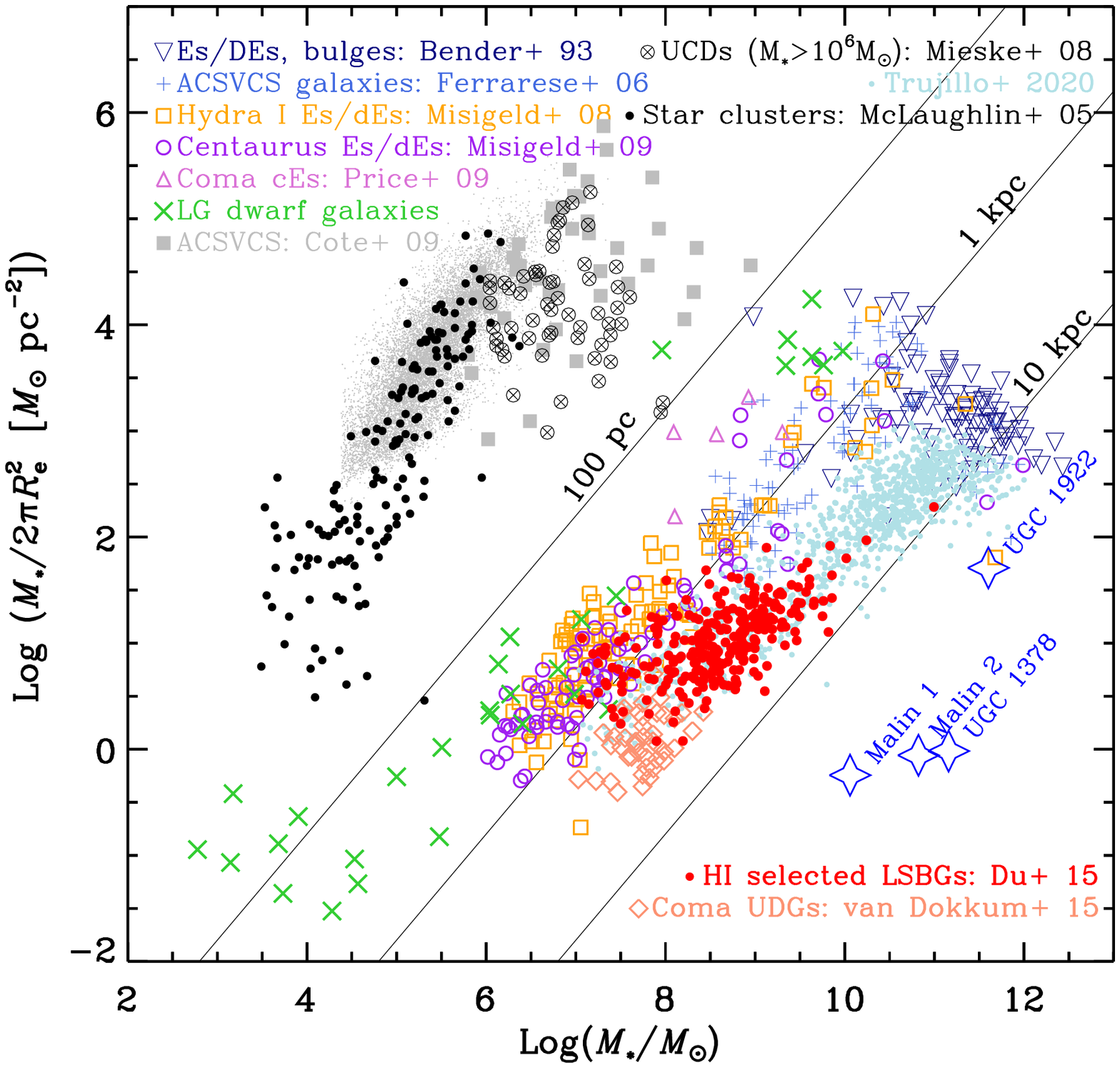}
    \caption{The stellar mass and mass density distribution of the different stellar systems. This figure includes the sample introduced in \citet{Misgeld2011}. We label the sample with reference. We add the sample of Coma UDGs \citep[][pink dimonds]{2015ApJ...798L..45V}, low redshift galaxies \citep[][small blue dots]{Trujillo2020} and the sample in this work (red filled circles). We also show the results of giant LSBGs such as Marlin 1 \citep{Malin1}, Marlin 2 \citep{Malin2}, UGC 1922 \citep{2018MNRAS.481.3534S}, UGC 1378 \citep{2019MNRAS.489.4669S} in open blue stars. We show the division line of the compact stellar system (thick line denoted by 100 pc), And the position of the stellar system with 1kpc, 10 kpc size. Our sample mainly has a similar radius to the massive galaxies and UDG sample. 
}
    \label{massdensity}
\end{figure}

\subsection{Size of the LSBG population}\label{massUVsize}

Fig. \ref{UVsize} and \ref{Hbandsize} show the size of the LSBGs in UV and H band. The consistent distribution of the H band mass size relation in Fig. \ref{Hbandsize} roughly indicates that the LSBGs are not exceptional in stellar mass distribution. However, the difference between the LSBG and other galaxies are clearly shown in Fig. \ref{UVsize} that LSBGs are typically 0.5 dex larger than the nearby galaxy sample. The magnitude and diameter limits of the GALEX NGS results yield an absent of the low surface brightness galaxy population \citep[see ][for more details]{2007ApJS..173..185G}. Therefore, our LSBG sample complements the GALEX NGS results, and show a larger radius than the local high surface brightness galaxies. 

The large scatter of the UV size, the stellar mass distribution may be caused by the complex dust extinction in massive galaxies, and the clumpy or irregular star formation morphology in low mass galaxies. Nevertheless, the LSBGs almost posses the largest UV size at a fixed stellar mass range. Previous studies show that the galaxy mass-size relation may correlate with the galaxy properties such as morphology, S\'ersic index, specific SFR, galaxy population and other factors\cite{2013MNRAS.434..325F, 2015MNRAS.447.2603L, 2017ApJ...838...19W}. We also show the galaxy morphology by the colour bar in Fig. \ref{UVsize}. Although the scatter of the UV size-stellar mass relation is large, galaxies with a similar morphology still show the trend that galaxies with the same morphology tend to have a tight correlation \citep{2013MNRAS.434..325F}. The location of the LSBGs in Fig. \ref{UVsize} implies a possible method to select LSBGs from the low mass, UV or U band disk galaxies. The current and future large field deep u band survey projects such as the CFHT large area U-band deep survey \citep[CLAUDS,][]{2019MNRAS.489.5202S} or the u band data released by Rubin Observatory \citep{2009arXiv0912.0201L, 2020arXiv200111067B} may be very helpful in selecting large LSBG sample efficiently.

\subsection{stellar mass build up of the LSBG}

Low mass galaxy sample in \citet{Trujillo2020} is selected from BOSS results \citep{2013MNRAS.435.2764M} which should not mainly consist by LSBGs. The consistency distribution in Fig. \ref{Hbandsize} indicate that the stars in LSBGs have already built as the other galaxies. However, Fig. \ref{UVsize} show that the star-forming size of the LSBGs is about 0.5 dex larger than the normal HSBGs, which is reasonable since the low surface brightness implies a larger radius. So we can expect the stellar mass would increase faster in the extended region of the LSBGs. 

We show the stellar mass and mass density ($\Sigma_{M_*} = M_*/2\pi R_{\rm H}^2$) of our LSBG sample in a big picture of the different stellar systems over 10 orders of magnitude in mass \citep{Misgeld2011, 2015ApJ...798L..45V, Eigenthaler2018} in Fig. \ref{massdensity}\footnote{The $R_{e}$ in Fig. \ref{massdensity} are measured from optical images. The difference of the galaxy sizes measured from H band and r band is less than 0.2 dex \citep[e.g., ][]{2012MNRAS.421.1007K}.}. The 100 pc line shows the division between the compact and extended stellar system \citep{Eigenthaler2018}. The LSBG systems locate in between the ultra-diffuse galaxies (UDGs) in Coma galaxy cluster and the massive galaxies, with about similar magnitude of the radius. Definition of the UDGs in recent studies are $R_e > 1.5$ kpc and $\left<\mu(r, R_e)\right> > 24$ mag / arcsec$^2$, which mainly have a lower surface brightness than our LSBG sample \citep[see the panel (d) of Fig. 3 in][]{Du2019}, and lower stellar mass. Both UDGs and LSBGs are diffuse galaxy population and have low star formation efficiency. And the adjacent location of the UDGs and LSBGs may imply that, if the UDS can have a sustained baryon inflow from IGM, and continuous star formation, then the UDGs may evolve to LSBGs, otherwise the UDGs may fade into a very diffuse, faint galaxy that below the detection limit. For the UGDs in the Coma cluster, the massive dark matter halo may sweep out the baryons nearby more efficiently, cut out the UGDs baryon supply, and quench or swallow the UDGs by ram pressure or merger \cite{2010ApJ...721..193P}. Therefore, the dense environment may cut down the connection between UDGs and LSBGs. One recent field UDG sample \citep{2020ApJS..247...46B} selected from the wide-field S-plus survey project \citep{2019MNRAS.489..241M} shows a typical stellar-mass about $10^8M_\odot$ and radius about 2.5 kpc, corresponding to about $\log(M_*/2\pi R_e^2)$ about 0.4, agree with the UDG properties from Coma galaxy cluster, implying that the origin of the diffuse nature in UDG may not only be caused by the environment.

For the stellar mass about $10^9M_\odot$, the HI selected LSBGs and low-z sample from \citet{Trujillo2020} in Fig. \ref{massdensity} have lower mass density than the elliptical galaxies, or the galaxy bulge, implying that the low mass sample in \citet{Trujillo2020} is mainly disk galaxies with no or weak central bulge structure. So the consistency of the stellar mass v.s. H band size relation also indicates that the stellar mass structure of the LSBGs is similar to the low mass disk galaxies.

The current star formation in the LSBGs would build stars at a larger radius (Fig. \ref{UVsize}), thus the LSBGs would evolve into an extended disk galaxy with higher stellar mass, such as the giant LSBGs such as Marlin 1 \citep{Malin1}, Marlin 2 \citep{Malin2}, UGC 1922 \citep{2018MNRAS.481.3534S}, UGC 1378 \citep{2019MNRAS.489.4669S} in Fig. \ref{massdensity}. Simulations of the low-z diffuse galaxy formation also show the stellar disk flatten caused by the supernova feedback \citep{DiCintio2017, DiCintio2019, Jiang2019}. Then the evolution trace of LSBGs would go toward the right lower direction in Fig. \ref{massdensity}, where there may be some Giant LSBGs. Upper panel of Fig. \ref{Hbandsize} also shows that galaxies with the stellar mass about $10^{10}M_\odot$ ($\sim 1$ dex higher than our sample) do not show a larger size. Thus we conclude that compaction might be necessary for the LSBGs \citep{Tacchella2016} to evolve into a massive high surface brightness galaxies.

\subsection{Size of the HI gas and the star formation activity}

\begin{figure}
    \centering
    \includegraphics[width=0.95\textwidth]{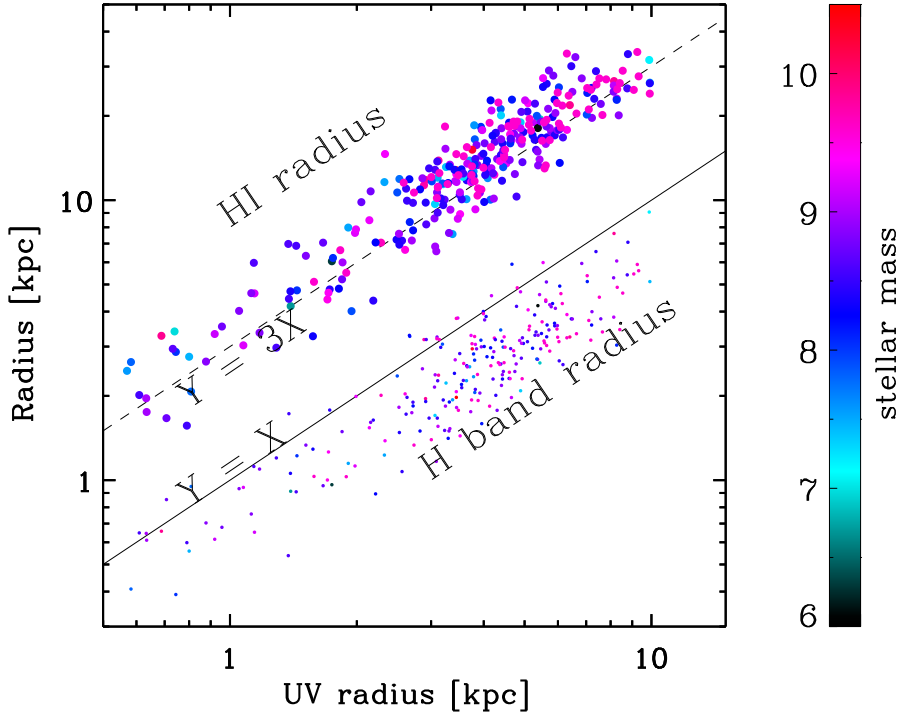}
    \caption{The HI gas radius, H band radius v.s. UV radii of our LSBG sample. The HI radius is estimated from the HI size-mass relation \citep{2016MNRAS.460.2143W}, which is a very tight correlation along 5 order of magnitude. We colour the sample by the stellar mass. The H band radius is 0.5 dex smaller than the UV radius, while the HI gas distributes at about 3 times larger radius.}
    \label{HIsize}
\end{figure}

Scale relation between the galaxy luminosity, surface brightness and thhe S\'ersic index show that the LSBGs tend to have flat morphology  \citep{2013pss6.book...91G, 2019PASA...36...35G}, which can be represented as a exponential disk. If the gas in LSBG follow the distribution: $\Sigma_{\rm gas} = \Sigma_0 e^{-r/r_s}$, where the half light radius is about $R_e = 1.678 r_s$ \citep{2010AJ....139.2097P}. So if we simply adopt Kennicutt-Schmidt law, the star formation disk would have $\Sigma_{\rm SFR} = A \Sigma_0 e^{-1.4 \times r/r_s} =  A \Sigma_0 e^{- r/(r_s/1.4)} $. Then the half light radius of star formation would be 1.4 times smaller than the gas radius.
Fig. \ref{HIsize} shows the size of UV and HI, which are approximately the SFR and gas radius, and about 3 times difference. So the Kennicutt-Schmidt law might have an index about 3 for LSBGs. Given the range of the gas surface density in our LSBGs sample is small \citep{Du2015}, the steep slope indicates that a small change of the gas surface density would alter the star formation surface brightness a lot. So the steep slope for LSBG would lead to a large scatter (or no) KS relation in LSBGs.

%
%
\subsubsection{Theoretical implication}

Theoretical works \citep{Mo1998, Amorisco2016, 2020MNRAS.491L..51P} and recent simulations show that the large size field galaxies may locate in the dark matter halo with high angular momentum \citep{Liao2019, 2019MNRAS.485..796M, Tremmel2019, Martin2019}. High rotation velocity will keep the baryons in a larger radius, and thus flatten stellar distribution.  Nevertheless, the formation of the H$_2$ from HI gas requires the high gas surface density and dust\citep{Leroy2008}. For our HI selected LSBG sample, the large size leads to a low HI surface density. So the LSBGs would still lack H$_2$. Therefore, from the theoretical point of view, galaxies with the large radii, such as LSBGs, should have a low star formation rate\citep{Lei2018}. Moreover, the high gas angular momentum would lead to a large gas size and even quench the galaxy by preventing the gas inflow \citep{2020MNRAS.491L..51P}. Therefore, if the large UV size of the LSBG is caused by the large gas angular momentum, we would expect the LSBGs quenched into red, extended galaxies, such as the red LSBGs revealed by HSC survey \citet{2018ApJ...866..112G}.  


\section{Conclusions}

We study the size of the HI-selected LSBG sample. Taking advantage of the low dust extinction in LSBGs, we make use of the UV image to represent the star formation distribution. We also use the H band image to trace the stellar mass distribution. We find the star-forming size of the LSBG is about 0.5 dex larger than the H band radius, so the star-forming activity in the LSBGs is still extended, rather than the compact star-forming cores \citep{Cheng2020}. The H band radius is consistent with the low-z disk galaxies, so the LSBGs may have a similar stellar structure, but much larger ongoing star formation region. The star-forming size and stellar mass of the LSBGs imply that the deep and wide-field blue band images may help us to select a large LSBG sample efficiently. The HI size of LSBGs is about 3 times larger than the UV, implying a steeper slope (or no) KS relation. We also discuss the possible link of the UDGs and our LSBGs and formation path of the LSBGs.

\begin{acknowledgements}
We thank the referee for carefully reading and constructive comments. C. C. is supported by the National Natural Science Foundation of China, No. 11803044, 11933003, 11673028 and by the National Key R\&D Program of China grant 2017YFA0402704. This work is sponsored (in part) by the Chinese Academy of Sciences (CAS), through a grant to the CAS South America Center for Astronomy (CASSACA). D.W. is also supported by the National Natural Science Foundation of China (NSFC) grant Nos. U1931109, 11733006, the Young Researcher Grant funded by National Astronomical Observatories, Chinese Academy of Sciences (CAS), and the Youth Innovation Promotion Association, CAS.
\end{acknowledgements}

\bibliographystyle{raa} 
\bibliography{ms2020-0224.bib}{} 

\end{document}